# Optimization of Bit Plane Combination for Efficient Digital Image Watermarking

Sushma Kejgir
Department of Electronics and Telecommunication Engg.
SGGS Institute of Engineering and Technology, Vishnupuri, Nanded, Maharashtra, India.
sbdabade@yahoo.co.in

Manesh Kokare
Department of Electronics and Telecommunication Engg.
SGGS Institute of Engineering and Technology, Vishnupuri, Nanded, Maharashtra, India.
mbkokare@sggs.ac.in

*Abstract:* **In view of the frequent multimedia data transfer authentication and protection of images has gained importance in today's world. In this paper we propose a new watermarking technique, based on bit plane, which enhances robustness and capacity of the watermark, as well as maintains transparency of the watermark and fidelity of the image. In the proposed technique, higher strength bit plane of digital signature watermark is embedded in to a significant bit plane of the original image. The combination of bit planes (image and watermark) selection is an important issue. Therefore, a mechanism is developed for appropriate bit plane selection. Ten different attacks are selected to test different alternatives. These attacks are given different weightings as appropriate to user requirement. A weighted correlation coefficient for retrieved watermark is estimated for each of the alternatives. Based on these estimated values optimal bit plane combination is identified for a given user requirement. The proposed method is found to be useful for authentication and to prove legal ownership. We observed better results by our proposed method in comparison with the previously reported work on pseudorandom watermark embedded in least significant bit (LSB) plane.**

I. INTRODUCTION:

*A. Motivation:*

Watermarking is an important protection and identification technique in which an invisible mark is hidden in the multimedia information such as audio, image, video, or text. It has been developed to protect digital signal (information) against illegal reproduction, modifications. The watermarking is also useful to prove legal ownership and authentication. A good fidelity transparent watermarking provides the watermark imperceptible to human visual system (HVS) that is human-eye cannot distinguish the original data from the watermarked data.

In the past literature on watermarking it is observed that bit plane method is one of the recommended methods of watermarking in spatial domain. This method is characterized by spread spectrum and is blind while watermark retrieval. Optimal implementation of this method maximizes the fidelity and robustness against different attacks. This method is based on the fact that the least significant bit plane of the image does not contain visually significant information. Therefore it can be easily replaced with watermark bits without affecting the quality of original image. However the survival of the watermark is an open issue and two main drawbacks of inserting watermark in least significant and most significant bits are:

- If watermark is inserted in least significant bit planes then the watermark may not survive against coding, channel noise, mild filtering or random bit-flipping.
- On the other hand, if the watermark is embedded in most significant bit plane, watermark survives but image quality is degraded.

Therefore, to get optimal results, in terms of fidelity, robustness, and high embedding capacity, a new bit plane modification method is proposed in this paper.

*B. Our Approach:*

To overcome above problems, we propose the novel method for image watermarking. Proposed method differs in two different ways than the earlier technique of bit plane watermarking. Firstly, to prove the ownership or identify the owner, most effective digital signature watermark is embedded instead of pseudorandom watermark. Secondly, instead of LSB, a previous bit to LSB is identified for watermark embedding to avoid the degradation of image and to survive the watermark after different general attacks like coding, channel noise, mild filtering or random bit-flipping. The advantages of the proposed method are summarized as follows.

- Proposed approach is optimal.
- Maximizes the fidelity.
- Maximizes the robustness against different attacks.
- Proposed method is having more payload capacity.

The rest of the paper is organized as follows: earlier related work to bit plane method is discussed in section 2. The proposed significant bit plane modification watermarking algorithm is discussed in section 3. The experimental results are presented in section 4, which is followed by conclusion and future scope in section 5.

## II. RELATED WORK

Sedaaghi and Yousefi [1] embedded the watermark in the LSB bit plane. In this method watermark is like a noise pattern i.e. pseudorandom pattern. The main disadvantage of this technique is that correlation coefficient (CRC) is very small. This shows that this method cannot withstand against attacks such as channel noise (small changes), bit flipping, etc. Yeh and Kuo [2] proposed bit plane manipulation of the LSB method and used quasi m-arrays instead of pseudorandom noise as a watermark. Here, watermark is recovered after the quantization and channel noise attacks. Gerhard et al. [3] discussed pseudorandom LSB watermarking, and highlighted the related work [4-8] where in LSB modifications are employed. They commented that LSB modification method is less robust and not much transparent.

In [9], two watermarking algorithms (LSB and discrete wavelet transform) are discussed by Xiao and Xiao. PSNR of LSB is reported to be higher i.e. 55.57 db. An experimental comparison for both against five attacks is made. LSB watermarking is reported to survive only against cropping. The simplest spatial domain image watermarking technique is to embed a watermark in the LSB of some randomly selected pixels [10]. The watermark is actually invisible to human eyes. However, the watermark can be easily destroyed if the watermarked image is low-pass filtered or JPEG compressed. In [11], advantages and disadvantages of LSB and most significant bit (MSB) watermarking are reported by Ren et al. To balance between robustness and fidelity, appropriate bit selection is proposed. Maeder and Planitz [12] demonstrated the utility of LSB watermarking for medical images. A comparison is also made with discrete wavelet transform based watermarking in terms of payload. Fei et al. [13] proposed MSB-LSB decomposition to overcome drawbacks of fragile authentication systems. However the use of LSB makes the system vulnerable to attacks. Kumsawat et al. [14] proposed the spread spectrum image watermarking algorithm using the discrete multiwavelet transform.

A threshold value is used for embedding the watermark strength to improve the visual quality of watermarked images and the robustness of the watermark. Chen and Leung [15] presented a technique for image watermarking based on chaos theory. Chaotic parameter modulation (CPM) is employed to modulate the copyright information into the bifurcating parameter of a chaotic system. Chaotic watermark is only a random bits, the problem of ownership identification is still unsolved. Cox et al. [16] advocated that a watermark should be constructed as an independent and identically distributed (i.i.d.) Gaussian random vector that is imperceptibly inserted in a spread-spectrum-like fashion into the perceptually most significant spectral components of the data. They argued that insertion of a watermark under this regime makes the watermark robust to signal processing operations (such as lossy compression, filtering, digital-analog and analog-digital conversion, re-quantization, etc.), and common geometric transformations (such as cropping, scaling, translation, and rotation) provided that the original image is available and that it can be successfully registered against the transformed watermarked image. Ghouti et al. [17] proposed a spread-spectrum communications watermark embedding scheme to achieve watermark robustness. The optimal bounds for the embedding capacity are derived using a statistical model for balanced multiwavelet coefficients of the host image. The statistical model is based on a generalized Gaussian distribution. BMW decomposition could be used constructively to achieve higher data-hiding Capacities. Bit error rate is graphically represented and not tested against geometric attacks.

## III. PROPOSED WATERMARKING ALGORITHM

Watermark embedding process and extraction process are shown in Fig. 1 and 2 respectively.

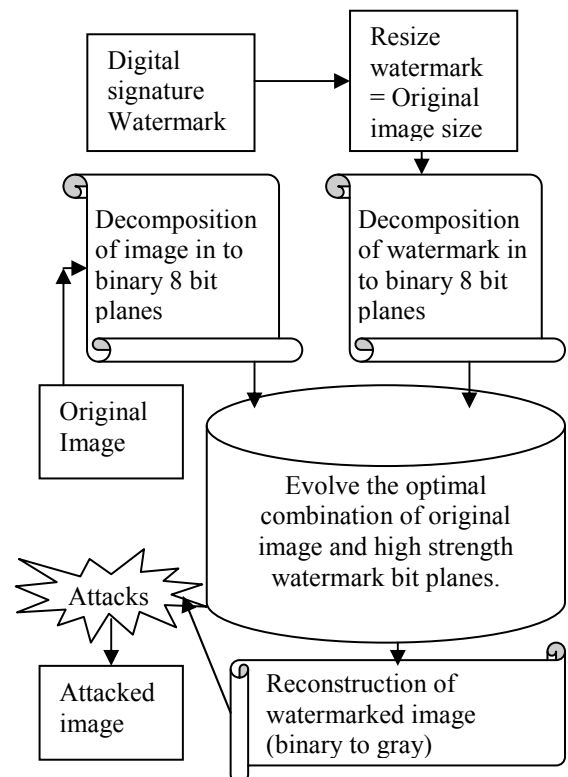

Fig. **1** Watermark embedding process

The proposed method is simple to apply, robust to different attacks and, has good fidelity to HVS. Broadly, in this method, original image and watermark are decomposed in to bit planes. Combinations of significant bit planes are searched to obtain optimal bit plane combination. Finally, using the identified bit planes watermarking is carried out.

*A. Watermark Embedding and Retrieval:*

In this proposed method, let $X(m,n)$ be the grey level image and $W(m,n)$ be the original digital signature watermark. The grey level image is transformed into the watermarked image $Y_W(m,n)$.

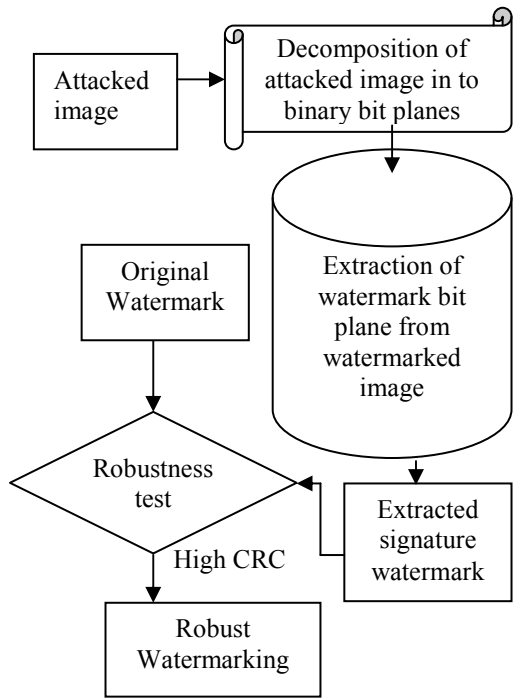

Figure **2** Watermark extraction process

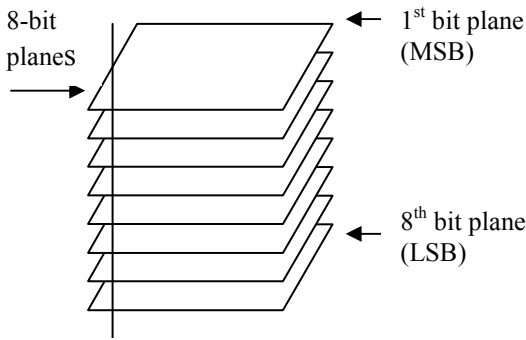

Fig. **3.** Bit plane representation of an image

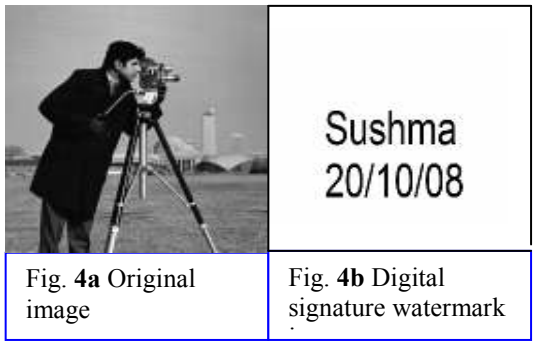

| Fig. **4a** Original image | Fig. **4b** Digital signature watermark |

The grey scale image $X$ is defined as follows: $X = \{X(m,n), m \in \{1,........,M\}, n \in \{1,........,N\}\}$, and $M, N$ are maximum dimensions of an image, where $X(m,n) \in \{0,........,255\}$ total number of grey levels.

Step by step algorithm for proposed method is explained below:

*Step 1: Decompose the grey level image to bit planes:* Grey level image is decomposed in to bit plane image. Each pixel in the image is represented by 8-bits. Therefore the image is decomposed into eight 1-bit planes, ranging from $8^{th}$ bit plane for LSB to $1^{st}$ bit plane for the MSB. The $8^{th}$ bit plane contains all the lowest order bits in the pixels comprising the image and $1^{st}$ bit plane contains all the higher order bits as shown in Fig. 3. Fig. 4a and 4b show grey level original image and digital signature watermark of dimension 256 x 256 respectively. These are decomposed in to bit planes as follows. Decomposition of original image in to 8-bit planes (refer Fig. 5):

$$X_l(m,n) = X_{b1}(m,n) + X_{b2}(m,n) + .......... + X_{b7}(m,n) + X_{b8}(m,n) \quad (1)$$

Similarly, decomposition of watermark in to 8-bit planes:

$$W_k(m,n) = W_{b1}(m,n) + W_{b2}(m,n) + .. + W_{b7}(m,n) + W_{b8}(m,n) \quad (2)$$

Where $l$ and $k$ indicates number of bit planes of image and $\in \{b1, b2,......b8\}$.

*Step 2: Replace the significant bit plane of original image with watermark bit plane:*

Following set of equations display replacement of $7^{th}$ bit plane of original image with $1^{st}$ bit plane of digital signature watermark as an example. The same procedure can be adopted for the remaining bit planes of the image.

$$Y_{b1}(m,n) = X_{b1}(m,n)$$
$$Y_{b2}(m,n) = X_{b2}(m,n)$$
-----
-----
-----
$$Y_{b7}(m,n) = W_{b1}(m,n)$$
$$Y_{b8}(m,n) = X_{b8}(m,n)$$

Resultant watermarked image is as follows:

$$Y_W(m,n) : Y_W(m,n) = Y_{b1}(m,n) + Y_{b2}(m,n) + ........ + Y_{b8}(m,n) \quad (3)$$

This bit plane watermarked image $Y_W(m,n)$ is recomposed in to grey level image $I(m,n)$.

*Step 3: Selection of significant bit planes of original image for watermarking:* Fig. 6 shows watermark embedded in all eight bit planes of original image by step 2. This is done so as to decide, by HVS, which bit planes of the image are good for watermarking. The bit plane, which does not degrade the image quality, after embedding watermark, is desirable. Accordingly the LSB ($8^{th}$ bit plane) and the one previous to LSB ($7^{th}$ bit plane) are most suitable as image quality is not degraded after watermark embedding. Therefore these bit planes shall provide good fidelity hence, selected for further analysis.

*Step 4: Formulation for watermarked image subjected to attacks:* In real life when watermarked image is distributed on the World Wide Web, it is encountered by different attacks. In this step, watermarked image is subjected to ten different types of attacks, leading to attacked image:

$$I_i^*(m,n), i \in \{1,2,....,10 \ different \ attacks\}$$

Extract the watermark bit plane from the attacked image. This retrieved watermark, after attack, is denoted as $W_{ibl}^*(m,n)$.

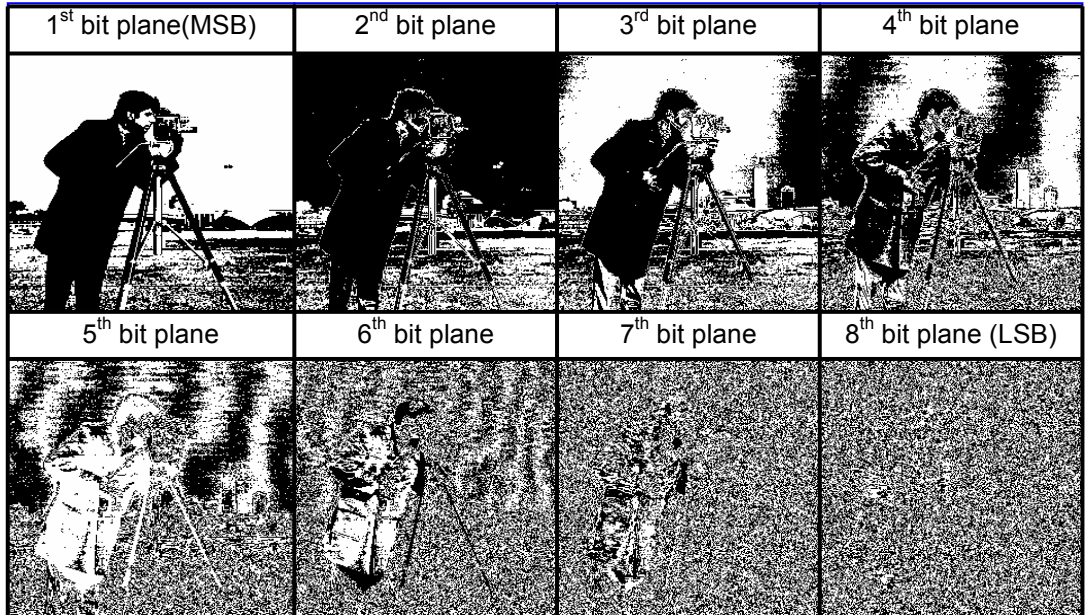

Fig. 5 Decomposed original image in to eight bit planes.

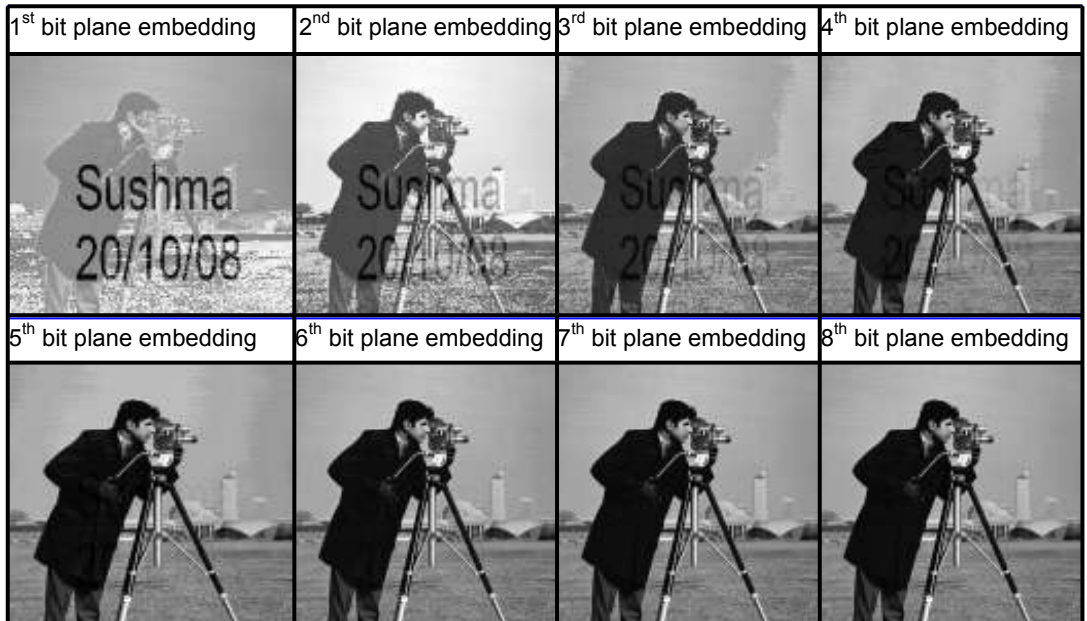

Fig. 6 Watermarked images after embedding watermark in all eight bit planes of image

Attacked image $I_i^*(m,n)$:

$$I_i^*(m,n) = I_1^*(m,n), I_2^*(m,n), .........I_i^*(m,n) \quad (4)$$

*Step 5: Watermark Retrieval:* In this step attacked image $I_i^*(m,n)$ is again transformed in to binary image i. e. 8-bit planes as shown below.

$$I_{il}^*(m,n) = I_{ib1}^*(m,n) + I_{ib2}^*(m,n) + ............ + I_{ib8}^*(m,n) \quad (5)$$

*Step 6: Computation of CRC:* Correlation coefficient between retrieved watermark and original watermark is estimated using a standard equation (6). The estimated correlation coefficients are denoted as $CRC_i \ (l,k)$. Where, I indicate different attacks, *l* is taken as 7th and 8th bit planes of original image as selected in step 3 and *k* denotes the bit planes of watermark from 1 to 8. The quality of watermarked image is observed by HVS. CRC varies between 0 and 1. CRC is defined as given below:

$$CRC = \frac{\sum_{n=1}^{256}\sum_{m=1}^{256} W(m,n) \times W^*(m,n)}{\sqrt{\sum_{n=1}^{256}\sum_{m=1}^{256} W(m,n) \sum_{n=1}^{256}\sum_{m=1}^{256} W^*(m,n)}} \quad (6)$$

$$if \ CRC_i(l,k) = \begin{cases} 0, & less \ robust \ watermarking \\ 1, & highly \ robust \ watermarking \end{cases} \quad (7)$$

$$PSNR = 10 \log_{10} \frac{(255)^2}{MSE} (db) \quad (8)$$

Mean square error is defined as:

$$MSE = \frac{1}{(m \times n)} \sum_{n=1}^{256}\sum_{m=1}^{256} (W(m,n) - W^*(m,n))^2 \quad (9)$$

| Watermarked image (8th bit image-8th pseudorandom watermark) | Original pseudo-random watermark | 1. Retrieval of watermark after angle rotation attack | 2. Retrieval of watermark after rotate transform attack |
|---|---|---|---|
| 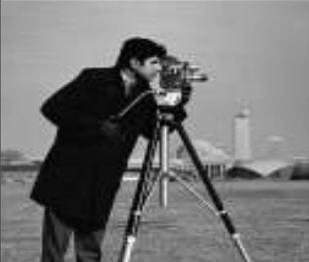 | 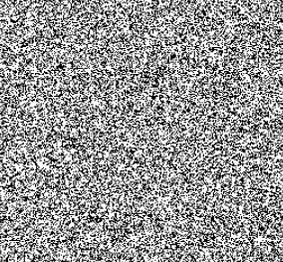 | 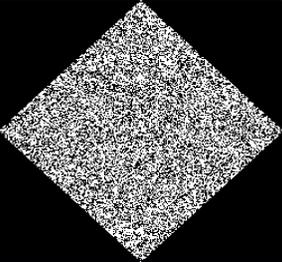 | 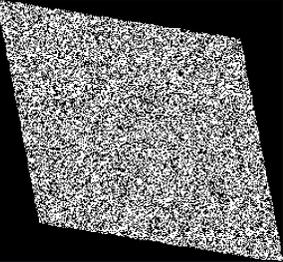 |
| 3. Retrieval of watermark after cropping of 41% attack | 4. Retrieval of watermark after low pass filter attack | 5. Retrieval of watermark after quantization attack | 6. Retrieval of watermark after translation motion attack |
| 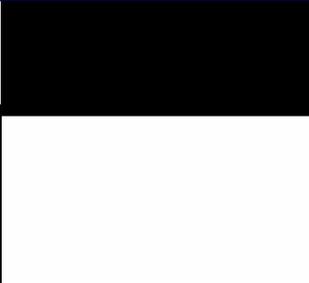 | 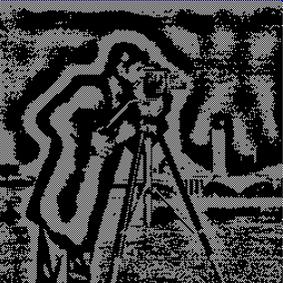 | 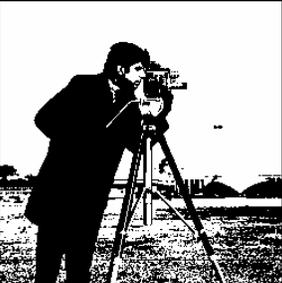 | 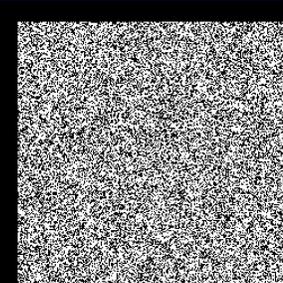 |
| 7. Retrieval of watermark after contrast stretching attack | 8 Retrieval of watermark after salt pepper attack | 9. Retrieval of watermark after compression attack | 10. Retrieval of watermark after shrinking attack |
| 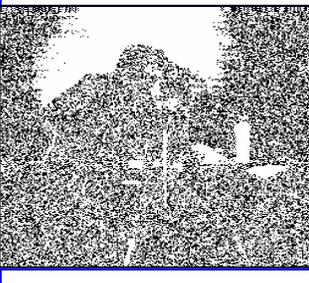 | 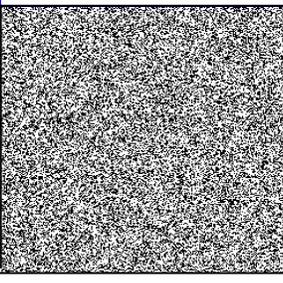 | 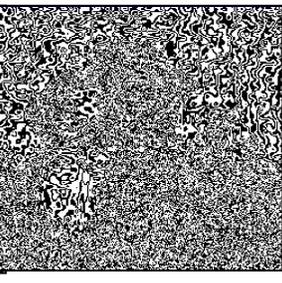 | 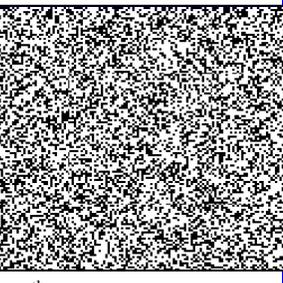 |

Fig 7. Result of retrieved watermarks after different attacks for **Existing method** (8th bit plane of original image replaced with 8th bit plane of pseudorandom watermark)

*Step 7: Estimation of peak signal to noise ratio (PSNR):* PSNR is calculated by using following equation. Capacity of the original image to carry the watermark is computed by measuring PSNR, which is defined as follows:

Where $W(m,n)$ is the original watermark, $W^*(m,n)$ is the extracted watermark after attack.

*Step 8: Weighted correlation coefficient computation:*

Weighted correlation coefficient is defined as follows:

$$Wt.\ CRC\ (l,k) = \sum_{i=1}^{10} CRC_i(l,k) \times a_i \qquad (10)$$

image and watermark under consideration. The step is repeated for combinations of selected bit planes of image and the entire bit planes of watermark respectively.

| Watermarked Image (7th bit image-1st digital signature watermark ) | Original digital signature watermark | 1. Retrieval of watermark after angle rotation attack | 2. Retrieval of watermark after rotate transform attack |
|---|---|---|---|
| 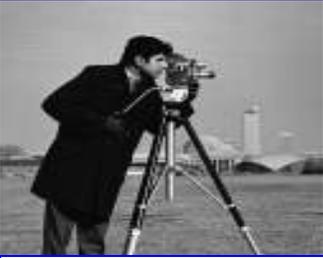 | 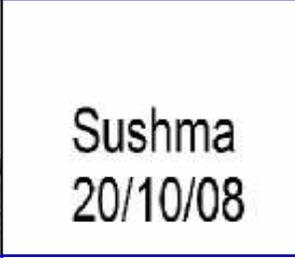 | 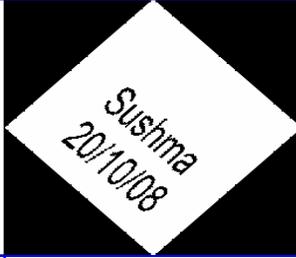 | 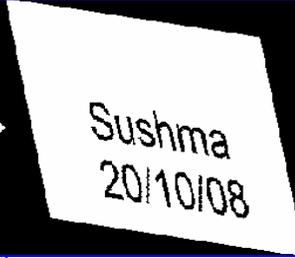 |
| 3. Retrieval of watermark after cropping of 41% attack | 4. Retrieval of watermark after low pass filter attack | 5. Retrieval of watermark after quantization attack | 6. Retrieval of watermark after translation motion attack |
| 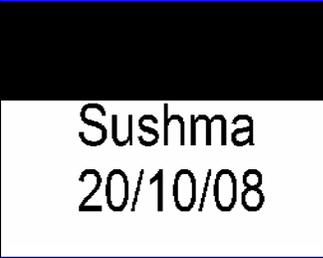 | 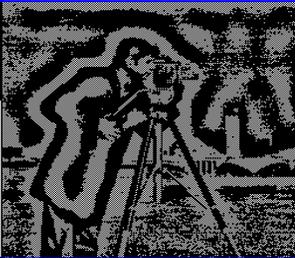 | 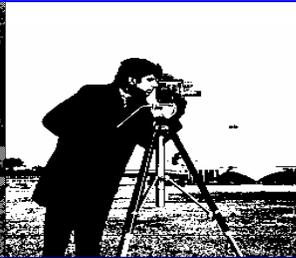 | 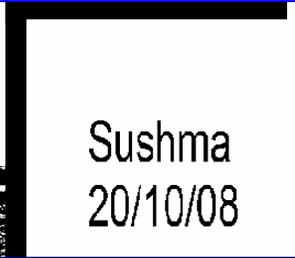 |
| 7. Retrieval of watermark after contrast stretching attack | 8 Retrieval of watermark after salt pepper attack | 9. Retrieval of watermark after compression attack | 10. Retrieval of watermark after shrinking attack |
| 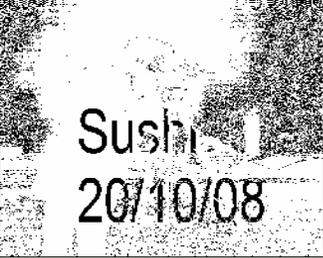 | 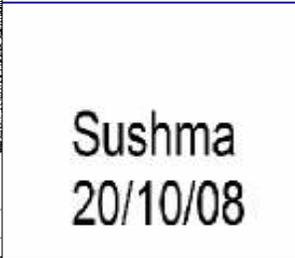 | 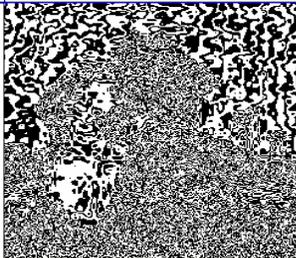 | 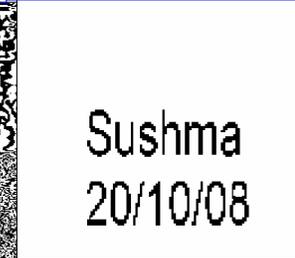 |

Fig 8. Result of retrieved watermarks after different attacks, for **proposed method** (7th bit plane of original image replaced with 1th bit plane of digital signature watermark).

Where, $a_i$ are the different weightings of attacks such that total $a_i = a_1 + a_2 + - - - - + a_{10} = 1$, and $i$ is the number of attacks. The identified attacks are assigned weightings based on damage caused, frequency, intensity and criticality or any other such criterion by the user. Based on these weightings, considering all the ten attacks, weighted correlation coefficient are estimated, for each bit plane combination of

*Step 9: Optimization:* The above step 8 is repeated by varying the weightings of attacks. The bit plane combination of original image and watermark for which, the weighted correlation coefficient is maximum, is selected as the optimized one for the given user requirements. This combination is used for optimized watermarking in terms of robustness and fidelity.

## IV. EXPERIMENTAL RESULTS

We have implemented our method on still grey scale image (dimension $256 \times 256$). In the subsections to follow extensive analysis is carried out to evolve the optimal combination of bit planes (image and watermark) to achieve desirable properties after watermarking.

For comparison original watermark is presented for each combination of the bit planes. Through these figures fidelity of watermarked image and survival of watermark after different attacks can be visually checked for the various combinations of image and watermark bit plane. Fig. 7 displays the results of existing method [1] (LSB, pseudorandom watermark embedding).

| Watermarked Image (8th bit image-8th digital signature watermark) | Original digital signature watermark | 1. Retrieval of watermark after angle rotation attack | 2. Retrieval of watermark after rotate transform attack |
|---|---|---|---|
| 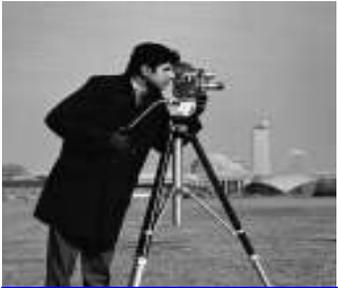 | 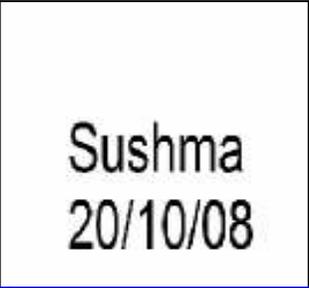 | 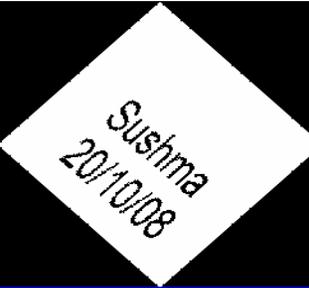 | 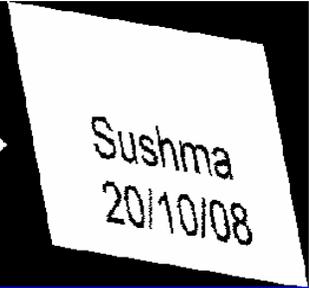 |
| 3. Retrieval of watermark after cropping of 41% attack | 4. Retrieval of watermark after low pass filter attack | 5. Retrieval of watermark after quantization attack | 6. Retrieval of watermark after translation motion attack |
| 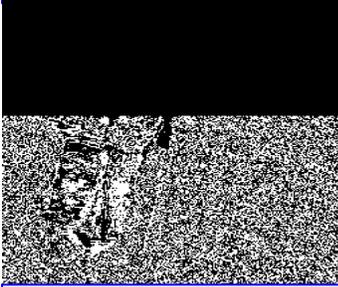 | 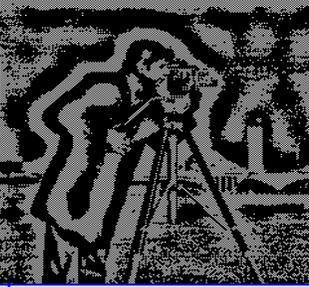 | 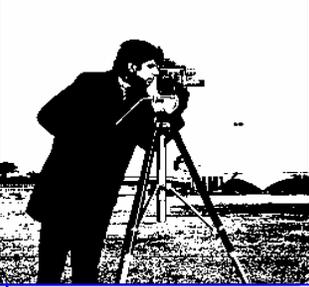 | 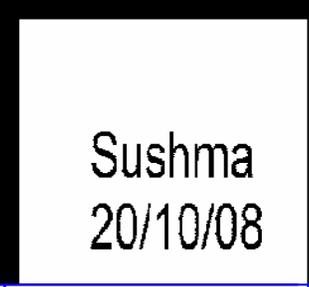 |
| 7. Retrieval of watermark after contrast stretching attack | 8 Retrieval of watermark after salt -pepper attack | 9. Retrieval of watermark after compression attack | 10. Retrieval of watermark after shrinking attack |
| 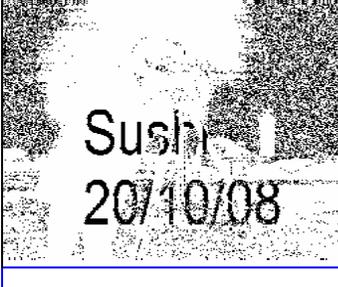 | 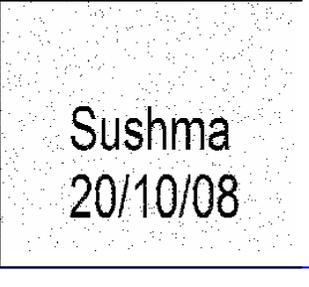 | 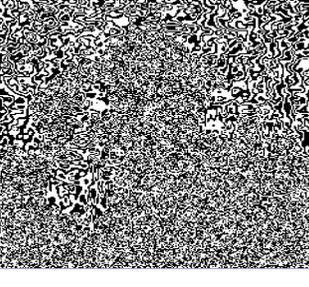 | 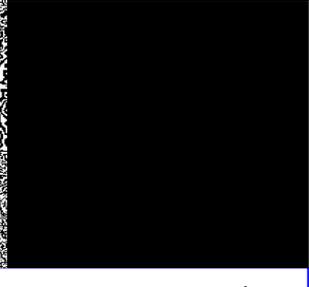 |

Fig 9. Result of retrieved watermarks after different attacks, for comparison **purpose to proposed method** (8th bit plane of original image replaced with 8th bit plane of digital signature watermark).

### A. Fidelity Checked by HVS

The results are displayed in Fig. 7, 8, 9, and 10. Each of these figures display watermarked image, original watermark, and retrieved watermark after different attacks. All these figures exhibit fidelity of watermarked image and survival of watermark after attacks.

Here, the watermark survives after seven different types of attacks out of ten. The retrieved watermark visually appears same as the original watermark, but automated correlation coefficient (standard method) is very small. This indicates that retrieved watermark is not similar to the original watermark. Fig. 8 shows the result of the combination, 1st bit plane of watermark embedded in 7th bit plane of original image, which shows survival of watermark against seven different attacks with good fidelity of watermarked image. Fig. 9 shows the results for other combination of bit planes (for example, 8th bit plane digital signature watermark embedded in 8th bit plane of original image). This result shows good fidelity but watermark survival is for minimum number of (five) attacks. Fig. 10 shows survival of watermark is good but fidelity of watermarked image is bad (1st bit plane watermark embedded in 1st bit plane of original image).

### B. CRC Results

CRC after different attacks and different combinations of bit planes is compared in Fig. 11. In this, CRC is plotted on y axis and different attacks are plotted on x axis as per the numbers is as follows:
1. Angle rotation attack.   2. Rotate Transform attack.
3. Crop attack 41%.   4. LPF (low pass filter) attack
5. Quantization attack.   6. Translation motion attack.
7. Contrast stretching attack. 8. Salt pepper attack.
9. Compression attack.   10. Shrinking attack.

| Watermarked Image (1st bit image-1st digital signature watermark ) | Original digital signature watermark | 1. Retrieval of watermark after angle rotation attack | 2. Retrieval of watermark after rotate transform attack |
|---|---|---|---|
| 3. Retrieval of watermark after cropping of 41% attack | 4. Retrieval of watermark after low pass filter attack | 5. Retrieval of watermark after quantization attack | 6. Retrieval of watermark after translation motion attack |
| 7. Retrieval of watermark after contrast stretching attack | 8 Retrieval of watermark after salt pepper attack | 9. Retrieval of watermark after compression attack | 10. Retrieval of watermark after shrinking attack |

Fig 10. Result of retrieved watermarks after different attacks, for comparison **purpose to proposed method** (1th bit plane of original image replaced with 1th bit plane of digital signature watermark).

Thus, above results indicate that the bit planes combination, i.e. 1st bit plane of watermark embedded in 7th bit plane of original image exhibit superiority over all other with respect to fidelity hence recommended by the proposed method.

CRC for different methods after different attacks, as given in legend: pseudo 8-8 indicates pseudorandom watermark (8th bit plane embedded in 8th bit plane), pseudo 1-1 indicates pseudorandom watermark (1st bit plane embedded in 1st bit plane), Signature 8-8 indicates digital signature watermark (8th bit plane embedded in 8th bit plane), etc. Graph shows that, for pseudo 8-8 and pseudo 1-1, CRC is nearer to the zero line, maximum CRC for combination of signature 1-1 but fidelity is bad for this method. The graph also shows that CRC is at higher level for the combination recommended by proposed method (signature 7-1). Also, for this fidelity is good as displayed in Fig. 8.

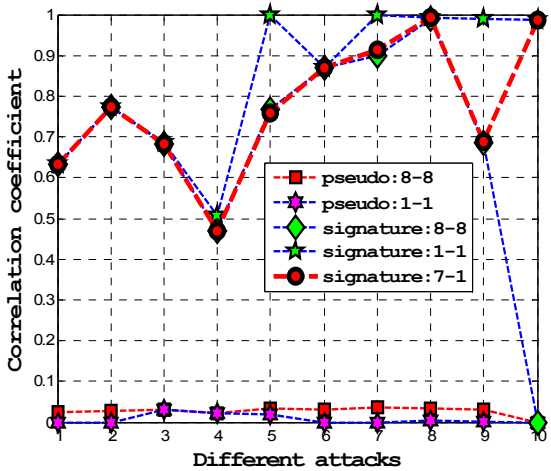

Fig 11. CRC for different attacks

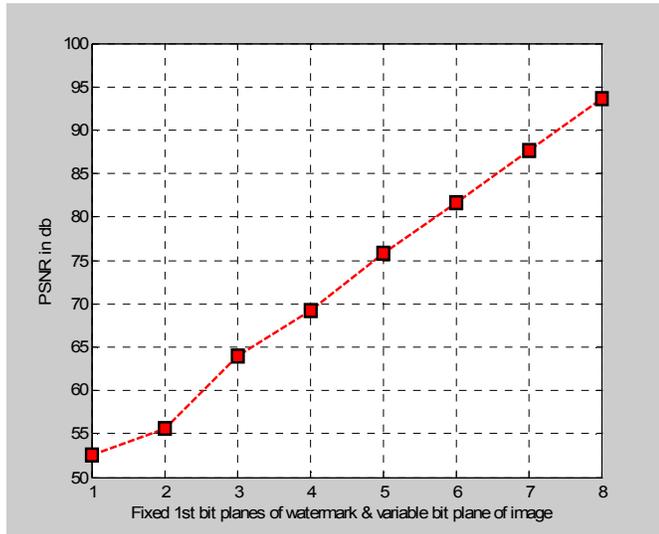

Fig. 12. PSNR for combinations of different bit planes of original image to 1st bit plane of watermark.

*C. PSNR Result*

In addition to above, for proposed bit plane combination watermark embedding capacity i.e. PSNR is observed to be high (87 db). PSNR for combinations of different bit planes of original image and 1st bit plane

(higher strength) of watermark is displayed in Fig. 12. From this it is observed that combination of (8th bit plane of image and 1st bit plane of watermark) is capable for higher pay load, but this combination is sensitive to small changes like bit flipping and robust to less number of attacks (refer Fig. 8 , 9 and table 1). Therefore previous bit plane (7th) of image is good for watermarking.

Table 1. Weighted CRC for different combination of original image and watermark bit planes are given.

| User requirements<br>Bit plane Combination : com.(l, k) | 1.Wt.CRC equal weights for all attacks i. e. $a_1$ to $a_{10}$=0.1 | 2.Wt.CRC $a_1$=0.05 $a_2$=0.05 $a_3$=0.05 $a_4$=0.05 $a_5$=0.05 $a_6$=0.05 $a_7$=0.2 $a_8$=0.2 $a_9$=0.2 $a_{10}$=0.1 | 3.Wt.CRC $a_1$=0.025 $a_2$=0.05 $a_3$=0.025 $a_4$=0.025 $a_5$=0.025 $a_6$=0.05 $a_7$=0.1 $a_8$=0.4 $a_9$=0.1 $a_{10}$=0.2 | 4.Wt.CRC $a_1$=0.025 $a_2$=0.025 $a_3$=0.05 $a_4$=0.05 $a_5$=0.05 $a_6$=0.05 $a_7$=0.05 $a_8$=0.2 $a_9$=0.3 $a_{10}$=0.2 |
|---|---|---|---|---|
| Com. (7,8) | 0.7854 | 0.8703 | 0.9212 | 0.8955 |
| Com. (7,7) | 0.7855 | 0.8703 | 0.9212 | 0.8955 |
| Com. (7,6) | 0.7857 | 0.8708 | 0.9220 | 0.8959 |
| Com. (7,5) | 0.7859 | 0.8713 | 0.9225 | 0.8962 |
| Com. (7,4) | 0.8106 | 0.8849 | 0.9285 | 0.9082 |
| Com. (7,3) | 0.8107 | 0.8849 | 0.9284 | 0.9081 |
| Com. (7,2) | 0.8110 | 0.8854 | 0.9291 | 0.9086 |
| Com. (7,1) | **0.8115** | **0.8861** | **0.9300** | **0.9091** |
| Com.(8,8) | 0.7855 | 0.8704 | 0.9213 | 0.8955 |
| Com. (8,7) | 0.7850 | 0.8700 | 0.9206 | 0.8950 |
| Com. (8,6) | 0.7854 | 0.8705 | 0.9215 | 0.8956 |
| Com. (8,5) | 0.7859 | 0.8712 | 0.9224 | 0.8962 |
| Com. (8,4) | 0.8108 | 0.8850 | 0.9286 | 0.9083 |
| Com. (8,3) | 0.8103 | 0.8847 | 0.9281 | 0.9078 |
| Com. (8,2) | 0.8108 | 0.8852 | 0.9289 | 0.9084 |
| Com. (8,1) | 0.8114 | 0.8859 | 0.9297 | 0.9090 |

*D. Weighted CRC Results*

In table 1, first column represents different bit plane combinations attempted in this work for digital image watermarking. Second column onwards represent results of weighted CRC, for different combinations, by varying the weightings of attacks. Here $a_1, a_2, ---, a_{10}$ represents different weightings of attacks respectively.

From results shown in table 1, it can be observed that the proposed method (1st bit plane of signature watermark embedded in 7th bit plane original image) provides the optimal combination yielding highest values of CRC as highlighted in the table. The table 1 highlights, optimal bit plane method which shows maximum robustness, in terms of CRC, for given user requirement.

## V. CONCLUSION

We observed that in previous bit plane methods survival of watermark appears to be good but CRC is nearer to zero level. The proposed method has the ability to perform better than the existing methods, based on bit plane, as higher CRC values are achieved. Also, when pseudorandom watermark is replaced with digital signature watermark there is rise in CRC indicating robustness of watermark. We observed that, in the image, the bit plane prior to LSB also does not contain visually significant information so it can be selectively used to optimally embed the watermark. Referring to results shown in Table 1, it can be concluded that proposed method leads to robust watermarking against geometric attacks and also yields highest correlation coefficient as compared to the previous bit plane method and other combination of bit planes. Also, it can be noted that PSNR value for proposed method is higher i. e. above 87 db.

It is noted that weighted correlation coefficient is useful to estimate the effect on the CRC, on account of change in user environment (in terms of variation in weight of the attack) while identifying the optimal bit plane combination. In future, the survival of watermark against various other different attacks can be checked.

**Sushma Kejgir** is an Assistant Professor of Department of Electronics and Telecommunication Engineering at Shri Guru Gobind Singhji Institute of Engineering and Technology, Vishnupuri, Nanded, India. Her subject of interest includes digital image watermarking and electromagnetic engineering.

**Dr. Manesh Kokare,** has completed his Ph.D. from the IIT, Kharagpur, India, in 2005. He is working as a faculty member in the Department of

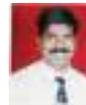

Electronics and Telecommunication Engineering at Shri Guru Gobind Singhji Institute of Engineering and Technology, Vishnupuri, Nanded, India. He has published about 35 papers in international and national journals and conferences. He received **Career Award for Young Teachers (CAYT)** for the year 2005 from AICTE, New Delhi, India. He is a life member of System Society of India, ISTE, and IETE and Member of IEEE, Member of IEEE Signal Processing Society, Member of IEEE Computer Society. He is a reviewer of fourteen international journals.